# A Stacked Autoencoder Application for Residential Load Curve Forecast and Peak Shaving

Xinan Wang, *Student Member, IEEE*, Yishen Wang, *Member, IEEE*, Di Shi, *Senior Member IEEE*,
Jianhui Wang, *Senior Member IEEE*,

*Abstract*— For the last ten years, utilities have observed on-going transitions on consumers' load curves. The previously flat load curves have more frequently turned into duck-shape. This is jointly caused by the increasing household loads as well as the popularity of rooftop solar photovoltaic. Such load curve transition challenges the operation flexibility of the existing systems and greatly increases the per-MWh energy costs. Peak shaving, in this context, becomes a critical task to demand-side management. Owing to the development of Battery Energy Storage Systems (BESS), numerous peak shaving strategies have been developed and implemented. In this paper, by applying a stacked autoencoder (SAE)-based residential peak load curve forecasting technology, we further lift the peaking shaving capabilities of BESSs to a new level. The proposed strategy takes into account the welfares of both generation-side and demand-side and reaches an optimal balance. A comprehensive case study using smart meter data demonstrates the effectiveness of the proposed method in peak shaving application.

## I. Background

A study conducted by the California ISO [1] shows the increasing penetration of rooftop solar photovoltaics (PVs), electric vehicles (EVs) and battery energy storage systems (BESS) is driving the net load curve to a duck-neck-shape. Such transition challenges the grid stabilities and increases the per-MWh energy costs. Therefore, demand-side management (DSM) becomes more important than ever.

As one of the most important DSM applications, peak shaving can help to mitigate such issue through demand response [2]-[5] and BESS [7]-[9]. This paper focuses on the BESS-based peak shaving methodology. Three common BESS-based strategies [7]-[9] are summarized in Figure 1. a-c. Figure 1.a is the full-output model, where a BESS supplies peak load until the battery is exhausted. The remaining peak load surge is supplied by the grid. Figure 1.b presents the threshold-based strategy [8], in which a BESS shaves the load when exceeding a pre-defined threshold. But the performance of the approach is heavily constrained by the limited BESS capacities. In Figure 1.c, BESS provides a conservative constant output to guarantee it can last longer than the peak-hour [9]. All these three strategies try to maximize consumers' welfares; on the other hand, they all neglect the utilities' welfares. Figure 1.d. shows an ideal peak shaving strategy, in which BESS' outputs dynamically follow the peak load curve with a constant mismatch. so that a flatten peak load curve is left to the utility which in other words significantly reduce the peak load volatility. Different from typical time series prediction works [10], the ideal peak-shaving requires a onetime full peak period curve prediction before peak-hour starts. The more accurate the prediction can be the higher peak can be shaved, our proposed SAE-based load curve prediction is detailed discussed in this paper to explain how it achieves a high prediction accuracy.

The biggest challenge for the ideal shaving is to predict an accurate peak load curve before peak-hour starts. Statistical-based load forecasting works [11]-[12] can hardly tackle the uncertainties brought by the residential BESS, PVs and EVs. In addition, feature-based supervised load prediction models like Artificial Neural Network (ANN), Support Vector Regression (SVR) [13]-[15] highly depend on the collected data, such as numerical weather index data and customer behavior analysis data, to reach satisfactory performances. And those data are not always available in practice. In this paper, we propose a machine learning-based load curve regression approach by using the unsupervised stacked autoencoder (SAE) algorithm. In this method, we use SAE to learn a representation of peak load curves from a training set and then encode the data into the SAE and decode it using the latent representation. In summary, the contributions of this paper are as follows:
- Strategically applies SAE to residential peak load forecasting and achieves a promising accuracy;
- Achieves the ideal peak shaving, which takes into account the welfares of both consumers' and utilities';
- A detailed performance analysis for the proposed framework is presented to serves as a reference for similar works in the field.

The remainder of this paper is organized as follows. In section II, an introduction of the proposed SAE is presented. In

X. Wang and J. Wang are with the Department of Electrical and Computer Engineering, Southern Methodist University, Dallas, TX 75205 USA (Email: xinanw@smu.edu; jianhui@smu.edu;). X. Wang, Y. Wang and D. Shi are with GEIRI North America, San Jose, CA 95134, USA.

This work is funded by SGCC Science and Technology Program under contract no. SGSDYT00FCJS1700676.

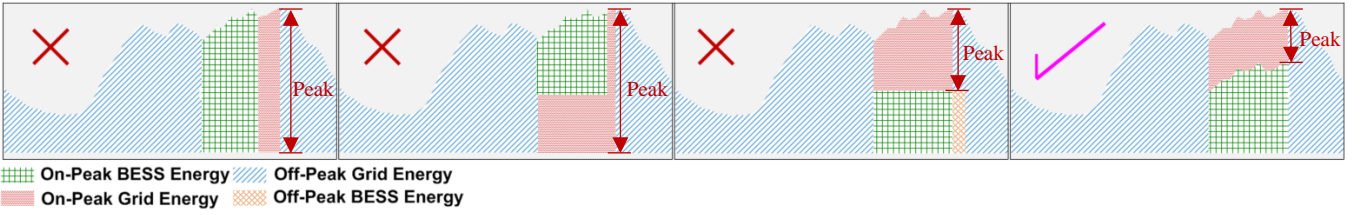

Figure 1.　(a). Full-output shaving　(b). Threshold-based shaving　(c). Constant-output shaving　(d). Ideal-shaving

battery storages peak shaving strategy. In section V, some section III, a detailed performance analysis of the proposed SAE is discussed. In section IV, case studies with real-world smart meter data demonstrate the effectiveness of residential concluding remarks are provided with discussions on the limitations of the presented framework and indicating future research works.

## II. INTRODUCTION OF STACKED AUTOENCODER

An autoencoder (AE) is one of the commonly used artificial neural network (ANN) structure in an unsupervised manner. It is widely adopted for data dimensionality reduction by learning a compressed representation $Z$ from the original data $X$ and reconstructing a $\tilde{X}$ which is close to $X$. Recently, some more powerful AEs such as sparse AE, stacked AE (SAE), convolutional AE (CAE), denoising AE (DAE) are proposed for image reconstruction, data noise removal, speech recognition and other machine learning applications.

### A. Structure of a SAE

A SAE is a stacked fashion of multiple shallow basic autoencoders, in which the output of one layer is the input of the successive layer. The training process starts from the first layer which encodes the input $X$ into a compressed hidden layer $Z^{(1)}$, and the decoder maps $Z^{(1)}$ back to $\tilde{X}$ so that the $\tilde{X}$ is close enough to $X$. When the first shallow autoencoder training is done, the decoder will be discarded. Then, $Z^{(1)}$ becomes the input of the second shallow autoencoder, which is used to train the further compressed hidden layer of $Z^{(2)}$. This greedy pre-training compression process ends until the last compressed hidden layer $Z^{(\frac{L-1}{2}+1)}$ (assumer $L$ is an odd number) is formed. The layer $Z^{(\frac{L-1}{2}+1)}$ is also called the bottleneck, since the decoder's structure of the SAE is a symmetry to the encoder's structure. The finally uncompressed output $Z^{(L)}$ has the same dimension as $Z^{(0)}$ or $X$. After a series of greedy pre-training are finished, a back-propagation based fine-tuning is conducted onto the entire SAE network so that the model is generic to all the training dataset. This training process of a SAE is shown in Figure 2. The mathematical formulation of such training process is presented as (1) and (2).

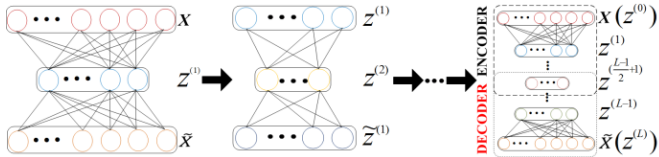

Figure 2.　Architecture of SAE

$$Z^i = \sigma^i(W^i Z^{i-1} + b^i), \forall\, i \in [1,2\cdots L], \quad (1)$$

$$\tilde{Z}^{i-1} = \sigma^{i'}(W^{i'} Z^i + b^{i'}), \forall\, i' \in [1,2\cdots L]. \quad (2)$$

Equation (1) is the encoding process of each layer; term $W^i$ is the weight matrix in the $i^{\text{th}}$ layer; $b^i$ is a constant bias vector in the $i^{\text{th}}$ layer; $\sigma^i$ is the non-linear activation functions of each compressing layer. The choices of activation functions typically include sigmoid, hyperbolic tangent and rectified linear unit (ReLU). Moreover, one can also custom an activation function to meet special needs as well. Equation (2) represents the decoding process of each layer, and the notations in (2) are similar to those in (1), but the parameters $W^{i'}$ and $b^{i'}$ are discarded after each hidden layer $Z^i$ is constructed. Equation (2) is only used to fine-tune the encoding part $W^i$ and $b^i$. The tuning process is in fact to minimize the reconstruction error between $Z^{i-1}$ and $\tilde{Z}^{i-1}$. The error can be written as $\mathcal{L}(Z^{i-1}, \tilde{Z}^{i-1})$. $\mathcal{L}$ is the loss function to evaluate the similarities between the input $Z^{i-1}$ and the output $\tilde{Z}^{i-1,K}$ ($K$ is the number of iterations), which can be cross-entropy loss, mean-square-error (MSE) or customized loss function when necessary.

### B. The Use of SAE

Ideally, an SAE is capable of extracting more non-linear features from the original data by stacking more layers, whereas PCA only projects the data into its linear subspace by maximizing variances [16]. Unlike lots of artificial neural networks, SAEs do not require hand-engineered features which are at risk of causing model performance degrading [17]. Instead, SAEs generate new features from each add-on hidden layers. Therefore, external negative influence such as erroneous features are isolated from the model training process. The training process of an SAE starts with the first layer AE, which maps the input $X$ into its compressed representative $Z^1$ through an encoder $\sigma^1$, then $Z^1$ is reconstructed into $\tilde{X}$ by a decoder $\sigma^{1'}$. Then, a fine-tuning process is done by back-propagation (BP) to minimize the differences between $X$ and $\tilde{X}$. After the e first AE is formed, the decoder $\sigma^{1'}$ is discarded and the hidden layer $Z^1$ becomes the input of next layer AE, similarly, a set of encoder $\sigma^2$, $Z^1$'s compressed representative $Z^2$ and decoder $\sigma^{2'}$ will be trained by the same p. This process will end until the last layer of SAE is formed. Then, a BP-based fine-tuning is conducted onto the whole model [18].

In this paper, we use SAE to forecast the missing data (also called masking noise [19]) because the compressed representation preserves the details of the missing part. Unlike a denoising autoencoder (DAE), SAE is not specially designed for noise removal (including masking noise). However, we proposed a novel SAE masking noise removal strategy which aims to outperform the DAE. A data with missing part can be represented using equation (3).

$$P_f(X_c|X) = diag(A) \cdot X + diag(I - A) \cdot C. \quad (3)$$

Where, $P_f$ is the process of corrupting the original data $X$ by masking noise. $X_c$ is the corrupted data. $A$ is a binary $n \times 1$ vector with $m$ ($m \leq n$) number of 1 and $n - m$ number of 0. $I$ is an $n \times 1$ identity vector. $C$ is a constant vector which represents the masking noise value. In $A$, the dimensions of 0 elements are the missing dimensions in $X$. The value of the constant vector $C$ can be arbitrary. When $C=[\mathbf{0}]$, the components of the noising dimensions are missing. Otherwise, the noising dimensions are masked by certain value.

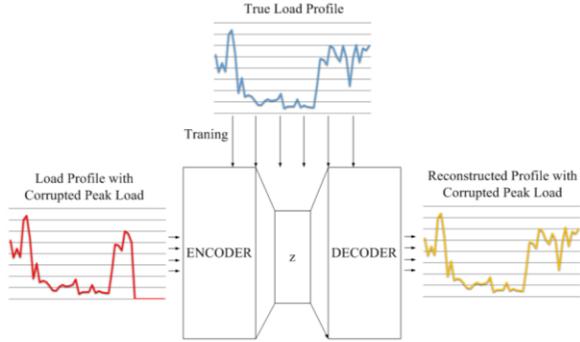

Figure 3. Peak load reconstruct process of SAE

Figure 3 shows the proposed SAE model, which is trained using clean data. The well-trained model can also fill the missing data in the input. This is inspired by the convolutional autoencoder (CAE)-based image reconstruction. A well-trained CAE can reconstruct the missing pixels in the same type of images [20]. In our work, we use SAE, which compresses data like a CAE but without using filters and pooling layers, to reconstruct the missing peak load curve. The challenge is that CAE reconstructs low-level pixel corruptions in an image [20], but the peak-hour duration counts for more than 20% of a day [21]. In this case, the prediction (filling) performance is sensitive to the value of $C$. By tuning the value of $C$, we can find the optimal point which yields the best performance, this will be discussed in detail in the following sections.

## III. SAE-BASED PEAK SHAVING METHODOLOGY

### A. Structures of The Proposed SAE

In [22], authors demonstrate that a SAE with more hidden layers can better capture the non-linear features in the input data. However, how many layers to choose is still data-dependent. In our case, we use the Irish consumers' smart meter readings with 30-min resolution, collected from the Commission for Energy Regulation (CER) [23]. The data contains 328 consumers' readings for 325 days. In this paper, we assume those 328 consumers live in the same community, and we conduct the peak shaving at such community level. In order to train our model with sufficient data, additional $C_{325}^2 = 52,650$ data is generated by taking the average of the unique combination of any two days' load curves. Together with the original 325 data, we have 52,975 data in total. Each data is a $48 \times 1$ vector, representing one day. By putting 45,000 data into the training set and the remaining 7,975 data into the out-of-sample testing set, we conduct the performances test of vanilla AE (3-layer), 5-layer, 7-layer and 9-layer SAE. The average testing result of 5-fold cross validation are shown in Figure 4.

In the test, we choose both mean square error (MSE) and mean absolute percentage error (MAPE) as the loss function. The structure for vanilla AE is 48-24-48. The latent structure for 5-layer SAE is 48-24-12-24-48. The latent structure for 7-layer SAE is 48-24-12-6-12-24-48. The latent structure for 9-layer SAE is 48-24-12-6-3-6-12-24-48. The results show that the 5-layer SAE suits our data the best. When choosing MSE as the loss function, the 5-layer SAE converges much faster than the 7-layer SAE and Vanilla AE. The converged RMSE is more than 1.4 kW lower than the rest. When choosing MAPE as the loss function, the 5-layer SAE converges at least 2 times faster than the rest. The converged MAPE is more than 0.76% lower than the rest.

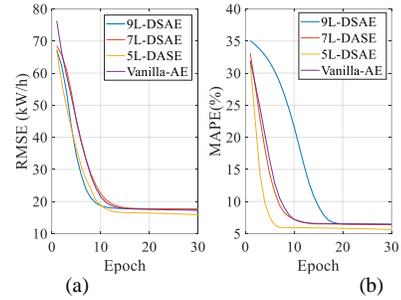

Figure 4. Performance comparison of different strcutures of AE using two loss function. (a) Loss function: MSE    (b)Loss function: MAPE

### B. Correct Weighting of Corrupted Dimensions

As discussed earlier, the input of the model is off-peak load curve with masked peak curve, and the output of the model is a full load curve. This means the model loss is measured on the entire curve using the conventional MSE-based loss function as shown in equation (4).

$$\mathcal{L}_2(X, \tilde{X}) = \frac{1}{n}\sum_{i=1}^{n}(X_i - \tilde{X}_i)^2. \quad (4)$$

Where $X_i$ is the $i^{th}$ component of the input $X$, $\tilde{X}_i$ is the $i^{th}$ component in the reconstructed output $\tilde{X}$. In fact, the model accuracy on off-peak part is of less interests. Rather than applying equal loss weight to every dimension in the reconstructed data, we put different weights on different portions of the data to improve peak load forecasting accuracy. The weight is assigned depending on the reconstruction difficulty or standard deviation of that portion of data. To achieve this goal, we designed a Weighted-MSE loss function which is shown in (5). The weight ratio is defined by (6) and (7), similar data variance-based weight assignment can be found in [24].

$$\mathcal{L}_{2,(\alpha,\beta)}(X, \tilde{X}) = \frac{\alpha}{n_c}\sum_{i \in \mathcal{C}(\tilde{X})}(X_i - \tilde{X}_i)^2 + \frac{\beta}{n_r}\sum_{i \notin \mathcal{C}(\tilde{X})}(X_i - \tilde{X}_i)^2, \quad (5)$$

$$\alpha + \beta = 1, \quad (0 \leq \alpha \leq 1, 0 \leq \beta \leq 1), \quad (6)$$

$$\frac{\alpha}{\beta} = \frac{n_r \cdot \sum_{i \in \mathcal{C}(\tilde{x})}\sqrt{\frac{\sum_{j=1}^{J}(X_{i,j} - \bar{X}_i)^2}{J-1}}}{n_c \cdot \sum_{i \notin \mathcal{C}(\tilde{x})}\sqrt{\frac{\sum_{j=1}^{J}(X_{i,j} - \bar{X}_i)^2}{J-1}}}. \quad (7)$$

Where, $\mathcal{C}(\tilde{X})$ denotes the index of the corrupted dimensions in $\tilde{X}$, $\alpha$ is the loss weight given to the corrupted dimensions, $\beta$ is the loss weight given to the non-corrupted dimensions, $n_c$ is the number of dimensions which are corrupted, $n_r$ is the number of the rest dimensions in the input or reconstructed output, and $J$ is the number of samples in the training dataset. By adjusting the value of $\alpha$ and $\beta$, one can put a variety of emphasizes on the model. For example, let $\alpha = 1$ and $\beta = 0$, then the accuracy of the reconstructed off-peak dimensions is neglected, the whole model is only tuned based on the accuracy of noised on-peak dimensions. Figure 5 shows the performance examples of two different weighted loss functions. In Figure 5.a, we make $\frac{\alpha}{\beta} = +\infty$, so the off-peak-hour data are neglected, only peak-hour data is considered. In Figure 5.b, we make $\frac{\alpha}{\beta} = 1$, so the off-peak-hour data has equally importance with the peak-hour data.

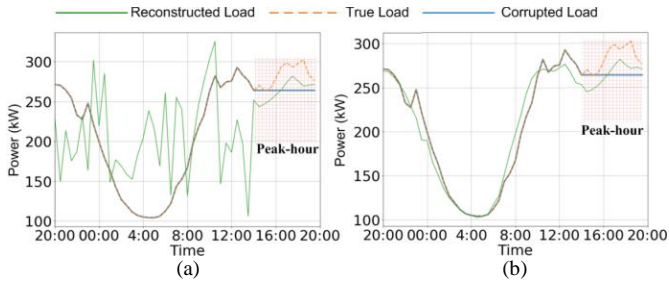

Figure 5. Weighted-MSE with varying $\frac{\alpha}{\beta}$ ratio. (a) $\frac{\alpha}{\beta} = \infty$ (b) $\frac{\alpha}{\beta} = 1$

We conduct a sensitivity analysis on 100 groups of tests. In each group, we train three models: 1) SAE with loss function of MSE; 2) SAE with loss function of the Weighted-MSE and $\frac{\alpha}{\beta}$ ratio calculated according to (7); 3) DAE. The performance validation is conducted using 7,975 testing data with the dimensions $X_{36} \sim X_{48}$ corrupted. The corruption level (value of $C$) for each group ranges from 0~1 p.u. (1 p.u. is the highest load in the dataset). The reconstructed peak-hour load curves are compared with the true data using RMSE.

The achieved RMSE for $C$ varying from 0~1 is shown in Figure 6 (a). It illustrates that DAE is insensitive to the masking noise value in $C$, and its RMSE stabilizes around 19.8 kW/h from corruption level 0-1 p.u. When masking noise is less than 0.5 p.u., DAE performs the best, but MSE based SAE is quickly approaching to DAE as masking noise increases. From 0.5~0.6, the performance of the weighted-MSE based SAE improves dramatically. It begins to lead the other two from 0.6~0.7 p.u. From 0.7~1.0 p.u., the DAE begins to lead again. Both Weighted-MSE and MSE reach their minimum RMSE, 17.11 kW/h and 20.20 kW/h, at masking noise value of 0.66. From this analysis, Weighted-MSE based SAE achieves the best performance when the masking noise value is 0.66. Its prediction RMSE outperforms the MSE based SAE by 3.10 kW/h and DAE by 2.57 kW/h.

We compared the performance of our method with an Artificial Neural Network (ANN) and an Extreme learning machine (ELM) using 5-fold cross validation. The testing result is shown in Table I. The results our proposed SAE-based method slightly outperforms the ELM as well as ANN.

TABLE I. PERFORMANCE COMPARISON

| | Peak Load Prediction RMSE (kW) | | |
|---|---|---|---|
| | ANN | ELM | SAE* |
| Fold I | 19.34 | 17.77 | 17.11 |
| Fold II | 20.21 | 17.98 | 16.74 |
| Fold III | 19.46 | 18.01 | 17.67 |
| Fold IV | 18.87 | 18.25 | 18.29 |
| Fold V | 19.29 | 18.27 | 17.35 |

The process of finding this optimal C value can be seen as a dictionary learning process, which is then used to conduct a load regression work. The similar dictionary learning process can be found in [25], in which the authors separate a mixed signal by learning the two signals' corresponding general deterministic dictionaries.

Figure 6 (b) explains the performance migration trend at different $\frac{\alpha}{\beta}$ ratio. The model performance reaches the best at the true $\frac{\alpha}{\beta}$ value which is 2.3. If the ratio is set too low, the performance curve is more flattened and the RMSE cannot dive down to a lower level. If the ratio is set too high, the whole curve will be lifted with the front end go downwards.

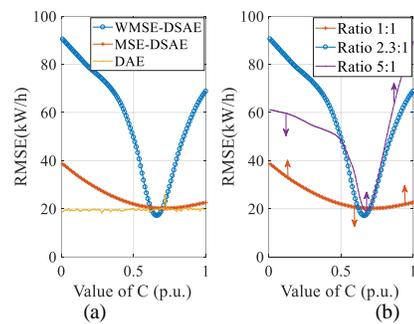

Figure 6. Performance comparison with different value of *C.* (a) between SAE and DAE. (b) between different $\frac{\alpha}{\beta}$ ratios.

IV. CASE STUDY

The effectiveness of the proposed SAE based peak load forecasting is demonstrated in Figure 7 with multiple examples.

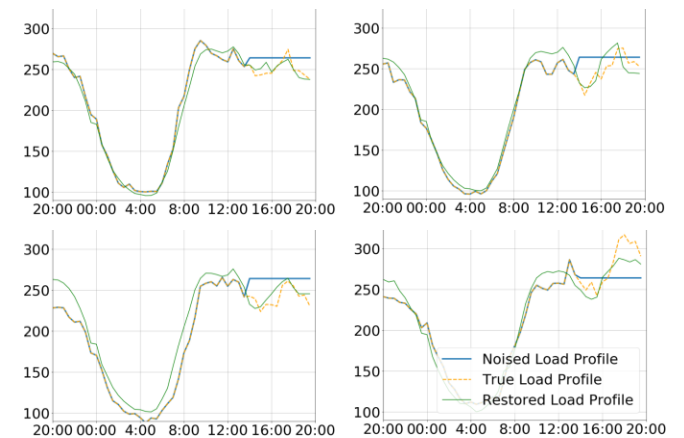

Figure 7. Peak load reconstruction cases

In Figure 7, presented examples demonstrate the peak load reconstruction capabilities of proposed SAE. Although some load curves are not perfectly matched, proposed SAE still captures the rise and fall trends well. In real-world applications, the BESS output power can be set according to the reconstructed curve. Table II provides a 10-day peak shaving level comparison using the 3 conventional methods shown in Figure 1.a-c and our proposed method. "A" refers to the full-output method, "B" refers to the threshold-based method, "C" refers to the constant-output method, and "D" refers to our proposed method. In this comparison, we assume one BESS with 500 kWh capacity is installed to provide peak shaving service for the community. According to energy usage pattern in the data, we define peak-hour lasts from 14:00-20:00. The battery is fully charged before peak-hour. The threshold for "B" is 150 kW, the constant output for "C" is set at the optimal point: 500/6=83.33kW.

TABLE II. PEAK SHAVING PERFORMANCE COMPARISION

| Day | Peak (kW) | Peak Shaving Level (%) | | | |
|---|---|---|---|---|---|
| | | A | B | C | D |
| Day1 | 260.77 | 0 | 11.50 | 31.96 | **35.31** |
| Day2 | 257.40 | 0 | 12.19 | 32.37 | **35.20** |
| Day3 | 302.65 | 0 | 0.62 | 27.53 | **29.99** |
| Day4 | 320.64 | 0 | 0 | 25.99 | **29.15** |
| Day5 | 325.14 | 0 | 0 | 25.63 | **26.89** |
| Day6 | 315.77 | 0 | 0 | 26.39 | **28.30** |
| Day7 | 301.70 | 0 | 1.07 | 27.62 | **28.11** |
| Day8 | 301.56 | 0 | 0 | 27.63 | **30.86** |
| Day9 | 300.01 | 0 | 0 | 27.78 | **33.82** |
| Day10 | 332.32 | 0 | 0 | 25.08 | **27.02** |

Table II demonstrates the effectiveness of our proposed SAE-based peak-shaving strategy. The full-output method can hardly provide any peak shaving due to the limited BESS capacity. The threshold-based method only works when the maximum peak occurs during the early peak-hour; otherwise, it is ineffective either. The constant-output method performs well but it is insensitive to the trend of the load. For consumers who need to pay for demand, this method cannot save them much. For our proposed method, it outperforms all the other methods. More importantly, it leaves a flattened load to the utilities which provides benefits to both consumers and utilities sides.

V. CONCLUSION

The proposed SAE-based peak shaving strategy can accurately forecast the peak load curves of residential consumers. The forecasted curve is used as the reference signal for BESS's output in peak-hour. The achieved peak shaving outperforms the conventional methods. The resultant peak-hour load maximizes the welfares of both the utilities and the customers. In the future work, we will develop a more sophisticated neural network which learns the possible optimal corruption level for each day, and the selected corruption level can be automatically attached to that day's load. In this way, we wish to achieve a higher load curve forecast accuracy.